\documentclass[12pt]{article}
\usepackage{amsmath}
\usepackage{amssymb}
\usepackage{xspace}
\usepackage{graphicx}               %better package for importing graphics
\usepackage{setspace}

\usepackage[usenames]{color}         %for colors

\newcommand{\mx}{\boldsymbol{x}}
\newcommand{\mX}{\boldsymbol{X}}
\newcommand{\me}{\boldsymbol{e}}
\newcommand{\my}{\boldsymbol{y}}

\newcommand{\mY}{\boldsymbol{Y}}
\newcommand{\mA}{\boldsymbol{A}}
\newcommand{\mB}{\boldsymbol{B}}
\newcommand{\mTau}{\boldsymbol{T}}
\newcommand{\mmu}{\boldsymbol{\mu}}
\newcommand{\malpha}{\boldsymbol{\alpha}}
\newcommand{\mSigma}{\boldsymbol{\Sigma}}
\newcommand{\mv}{\boldsymbol{v}}

\usepackage{hyperref}

\usepackage[authoryear, comma, sort&compress]{natbib}

\title{On optimal direction gibbs sampling}

\author{
J. Andr\'es Christen$^{(a)}$\thanks{Corresponding author.}, Colin Fox$^{(b)}$, \\
\medskip
Diego Andr\'es P\'erez-Ruiz$^{(a)}$ and Mario Santana-Cibrian$^{(a)}$ \\
(a) Centro de Investigaci\'on en Matem\'aticas (CIMAT), \\
 A.P. 402, Guanajuato, Gto. 36000, Mexico. \textit{jac@cimat.mx} \\
(b) Colin Fox, Department of Physics, University of Otago, \\
New Zealand. \textit{fox@physics.otago.ac.nz}
}

\date{May 2012}

\begin{document}

\bibliographystyle{spbasic}

\maketitle

\begin{abstract}
Generalized Gibbs kernels are those that may take any direction not necessarily bounded
each axis along the parameters of the objective function.  We study how to optimally choose
such directions in a Directional, random scan, Gibbs sampler setting.  The optimal direction
is chosen by minimizing to the mutual information (Kullback-Leibler divergence) of two
steps of the MCMC for a truncated Normal objective function.  The result is generalized
to be used when a Multivariate Normal (local) approximation is available for the objective
function.  Three Gibbs direction distributions are tested in highly skewed non-normal 
objective functions.
\end{abstract}

\section{Introduction}

Let $\pi(\mx)$ be the objective distribution of interest.  Let $\mX \in \mathbb{R}^n$ be a random variable with density $\pi(\mx)$.
The (univariate) Gibbs sampler is a MCMC sampling algorithm that simulates systematically or randomly from the conditional distributions
\begin{equation} \label{eqn.fullcond}
f_{ X_i | \mX_{-i} } ( x_i | \mx_{-i} ) \propto \pi(\mx) ,
\end{equation}
where the notation
\begin{equation}
\boldsymbol{v}_{-i} = ( v_1, \ldots , v_{i-1}, v_{i+1}, \ldots , v_n) \nonumber
\end{equation}
represents the $(n-1)$-dimension vector created by deleting the $i$-th entry from the $n$-dimension vector $\boldsymbol{v}$.  (\ref{eqn.fullcond})~above represent univariate distributions and are the conditional distributions alone the axis for the base chosen to represent $\mX$, the so called full conditional distributions.

\noindent Using (\ref{eqn.fullcond}) a Markov chain $\mX^{(1)}, \mX^{(2)}, \ldots $ is created considering the transition kernel
\begin{equation}
K_i( \mx^{(t)}, \mx^{(t+1)} ) = f_{ X_i | \mX_{-i} } ( x^{(t+1)}_i | \mx^{(t)}_{-i} ) 
  1( \mx^{(t+1)}_{-i} = \mx^{(t)}_{-i} ). \nonumber
\end{equation}
That is, the $i$-th kernel changes only the $i$-th coordinate by simulating from the full conditional distribution $f_{ X_i | \mX_{-i} } ( \cdot | \mx^{(t)}_{-i} ) $. In random scan Gibbs, a complete transition kernel is defined as
\begin{equation}
K( \mx, \my ) = \sum_{i=1}^n w_i K_i ( \mx, \my ) , \nonumber
\end{equation}
\noindent for some weights $\sum_{i=1}^n w_i = 1$; $w_i \geq 0$.

The direction Gibbs sampler generalizes this idea by choosing a direction $\me \in  \mathbb{R}^n$, $|| \me || = 1$, and sampling from the conditional distribution alone such direction. This can be written as
\begin{equation}
\mX^{(t+1)} = \mx^{(t)} + r \me , \nonumber
\end{equation}
where the length $r \in \mathbb{R}$ has distribution proportional to $\pi(\mx^{(t)} + r \me)$.  In any case, it can be seen that the transition kernel has detailed balance with respect to $\pi$, and by assuring $\pi$-irreducibility the Markov chain has $\pi$ as ergodic distribution.

The natural question to ask is how to choose $\me$ to optimize the convergence (mixing) of the Markov chain? Indeed, once irreducibility is assured, any chain will have the correct ergodic distribution ($\pi$) but performance will depend on how dependent $\mX^{(t+1)}$ and $\mX^{(t)}$ are. In this context a convenient, although less known, dependence measure to use is the \textit{mutual information} $I$ between random variables $X$ and $Y$, which measures the Kullback-Leibler divergence between the joint model $f_{X,Y}$ and the independent alternative $f_{X} f_{Y}$, that is, 
\begin{equation}
I( \mY, \mX ) = \int \int f_{ \mY, \mX } ( \my, \mx ) 
 \log \frac{f_{ \mY, \mX } ( \my, \mx )}{ f_{ \mY } ( \my ) f_{\mX } ( \mx )} d\mx d\my. \nonumber
\end{equation}
In our case, this translates to $ \mX = \mX ^{(t)}$ and $ \mY = \mX ^{(t+1)}$.
Assuming that $\mX \sim \pi$, we see that $f_{ \mY } ( \my ) = \pi( \my )$ and $f_{ \mY, \mX } ( \my, \mx ) = \pi(\mx) K(  \mx , \my ) $.  Therefore, the mutual information above can be calculated as
\begin{equation}\label{eqn.mut}
I( \mY, \mX ) = \int \int \pi(\mx) K(  \mx , \my ) \log \frac{K(  \mx , \my )}{\pi(\my)} d\my d\mx .
\end{equation}

The idea would be to choose directions for which $I( \mX^{(t+1)}, \mX^{(t)} )$ is minimized.  Since $I$ is a Kullback-Leibler divergence it is well defined, $I \geq 0$ and $I = 0$ if and only if $\mX^{(t+1)}$ and $\mX^{(t)}$ are independent ($f_{X,Y} = f_{X} f_{Y}$ a.s.). Finding directions $\me$ for which $I$ is minimum will provide our optimization criterion, to obtaining better suited direction Gibbs samplers.

\section{The Normal case}\label{sec.normal}

In many applications, a Multivariate Normal approximation to the posterior is possible.  We will try to use such Normal approximation to produce an improved Gibbs samplers. In this section we will assume $\pi$ to be Normal, a case that lends itself to calculate $I( \mX^{(t+1)}, \mX^{(t)} )$.
For the moment, we will consider that the dimension is sufficiently low so as to have the eigen decomposition of the precision matrix available.  

We therefore assume that $\pi$ is a multi- variate Normal with \textit{precision} ($n \times n$) matrix $\mA$ (the inverse of the variance-covariance matrix) and mean column vector $\mmu$.  As explained above, given a direction $\me \in  \mathbb{R}^n$, $ || \me || = 1$, the chain moves from $\mX^{(t)} = \mx$ to 
\begin{equation}
\mY = \mX^{(t+1)} = \mx + r \me , \nonumber
\end{equation}
where the length $r \in \mathbb{R}$ has distribution $g$ proportional to $\pi(\mx + r \me)$.  That is
\begin{equation}
g(r) \propto \exp \left\{ -\frac{1}{2} (\mv + r\me )' \mA (\mv + r\me ) \right\} , \nonumber
\end{equation}
where $\mv = \mx - \mmu$.  After some algebra we see that $r \sim N\left( -\frac{\me' A \mv}{\me' A \me}, \me' A \me \right)$, where $\me' A \me$ is the \textit{precision} (by setting $\me = \me_i$, the $i$-th standard bases vector, one obtains
\begin{equation}
Y_i \sim N \left( \mu_i - a_{ii}^{-1} (\mv' \mv - (x_i - \mu_i)^2) ,   a_{ii} \right) , \nonumber
\end{equation}
which is the full conditional distribution for a Multivariate Normal distribution for entry $ i $,
thus returning to the usual Gibbs sampler).  From this we see that the transition kernel corresponding to direction $ \me $ is
\small
\begin{equation}
K_{\me} ( \mx , \my ) = \left( \frac{\me' \mA \me}{ 2 \pi } \right)^{1 \over 2} 
 \exp \left\{ -\frac{\me' \mA \me}{2}  \left( \me' (\my - \mx) + \frac{\me' \mA \mv}{\me' \mA \me} \right)^2 \right\} 
 1( \my = \mx + \me' (\my-\mx) \me ) \nonumber
\end{equation}
\normalsize
(note that $\my - \mx = r \me$ and since $\me' \me = 1$, $r = \me' (\my - \mx)$; therefore $\my$ is restricted to the line $\my = \mx + \me' (\my-\mx) \me$).

We calculate $I_{\me}( \mX^{(t+1)}, \mX^{(t)} )$, the mutual information of the Gibbs sampler given direction $\me$, as in~\ref{eqn.mut}.  Note that
\begin{equation}
\log \frac{K_{\me} ( \mx , \my ) }{\pi(\my)} = C + \frac{1}{2} \log \me' \mA \me 
 - \frac{1}{2} \left( Q_1(\me, \mx, \my) - Q_2(\my) \right) , \nonumber
\end{equation}
where 
\begin{align}
C &= \frac{n-1}{2}\log 2\pi - \frac{1}{2} \log \left| \mA \right| , \nonumber \\
Q_1(\me, \mx, \my) &= \me' \mA \me \left( \me' (\my - \mx) + \frac{\me' \mA \mv}{\me' \mA \me} \right)^2 , \nonumber \\
Q_2(\my) &= (\my - \mmu)' \mA (\my - \mmu) . \nonumber 
\end{align}
From this we see that
\begin{equation}
\int \log \frac{K_{\me} ( \mx , \my ) }{\pi(\my)} K_{\me} ( \mx , \my ) d\my = 
C - \frac{1}{2} + \frac{1}{2} \log \me' \mA \me  + \frac{1}{2} \int Q_2(\my) K_{\me} ( \mx , \my ) d\my \nonumber
\end{equation}
since $\int Q_1(\me, \mx, \my) K_{\me} ( \mx , \my ) d\my = 1$. The integral $\int Q_2(\my) K_{\me} ( \mx , \my ) d\my$ may be calculated by transforming back to $r$ since
\small
\begin{equation}
\int Q_2(\my) K_{\me} ( \mx , \my ) d\my = \int (r\me - \mv)' \mA (r\me - \mv) g_{\me} (r) dr. \nonumber
\end{equation}
\normalsize
After some algebra one sees that
$$
\int Q_2(\my) K_{\me} ( \mx , \my ) d\my = 1- \frac{\mv' \mA \me \me' \mA \mv}{\me' \mA \me} + \mv' \mA \mv .
$$
Therefore
\begin{equation*}
\int \log \frac{K_{\me} ( \mx , \my ) }{\pi(\my)} K_{\me} ( \mx , \my ) d\my = 
C + \frac{1}{2} \log \me' \mA \me - \frac{1}{2}\frac{\mv' \mA \me \me' \mA \mv}{\me' \mA \me} + \mv' \mA \mv .
\end{equation*}
We need now to integrate with respect to $\pi(d\mx)$.  We note that $\int \mv' \mA \mv \pi(\mx) d\mx = n$.  Moreover, the expected value
of a quadratic form is
\begin{equation}
E(\boldsymbol{z}' \boldsymbol{R} \boldsymbol{z}) = tr(\boldsymbol{R} \boldsymbol{\Sigma} ) + \mmu' \boldsymbol{R} \mmu , \nonumber
\end{equation}
where $\mmu$ and $\boldsymbol{\Sigma}$ are the mean vector and the variance-covariance matrix of $\boldsymbol{z}$.
Letting $\boldsymbol{R} = \frac{\mA \me \me' \mA}{\me' \mA \me}$ and since $E(\mv) = \boldsymbol{0}$ we obtain
\begin{align}
E\left( \mv' \frac{ \mA \me \me' \mA}{\me' \mA \me} \mv \right) &= \frac{1}{\me' \mA \me} tr(\mA \me \me' \mA \mA^{-1}) \nonumber \\
&= \frac{1}{\me' \mA \me} tr(\mA \me \me' ) \nonumber \\
&= \frac{1}{\me' \mA \me} tr(\me \mA \me' ) \nonumber \\
&= 1. \nonumber
\end{align}
Therefore
\begin{align}\label{eqn.Ie}
I_{\me} ( \mX^{(t+1)}, \mX^{(t)} ) &= C + n - \frac{1}{2}  + \frac{1}{2} \log \me' \mA \me \nonumber \\
&= \quad C_1 + \frac{1}{2} \log \me' \mA \me ,
\end{align}
where $C_{1} = C + n - \frac{1}{2} $.

\subsection{Choosing a set of directions}\label{sec.directions}

We need now a distribution $h$ for directions to be chosen to generate an \textit{irreducible} Gibbs sampler.
That is, according to (\ref{eqn.Ie}) the best direction is that that minimizes $C_1 + \frac{1}{2} \log \me' \mA \me$, but
we cannot simply choose the best direction: the resulting Gibbs will not be irreducible and clearly we will not be
sampling from $\pi$.  The chain most be $\pi$-irreducible in order that the chain has as ergodic distribution $\pi$.
Indeed, if directions have distribution $h(\me)$, and this distribution has the whole sphere $\mathbb{S}^n$ as support, then
the resulting Markov chain $K( \mx , \my ) = \int K_{\me}( \mx , \my ) h(\me) d\me$ is irreducible.

\citet{Kaufman:Smith:94} argue that an optimal direction distribution is
\begin{equation}
h(\me) \propto \sup_{\mx \in \mathcal{X} , r \in \mathbb{R}}
\left\{ \int \pi( \mx + \tau\me ) d\tau \frac{| r |^{n-1}}{\pi( \mx + r\me )} \right\} , \nonumber
\end{equation}
as far as optimizing the (geometric) rate of convergence of the resulting Gibbs sampling.  However, this only applies
for a bounded support $\mathcal{X}$ for $\pi$.  Little else has been said regarding the optimal directions for generalized Gibbs
samplers.  We cannot control in general the term $\frac{| r |^{n-1}}{\pi( \mx + r\me )}$ for unbounded support.
However, this suggest choosing a direction 
\begin{equation}
h^*(\me) \propto \sup_{\mx \in \mathcal{R}^n } \left\{ \int \pi( \mx + \tau\me ) d\tau  \right\} . \nonumber
\end{equation} 
For the normal case, it is not difficult to see that
\small
\begin{equation}
\int \pi( \mx + \tau\me ) d\tau \leq \frac{(2\pi)^{-\frac{n-1}{2}} | \mA |^{\frac{1}{2}}}{\sqrt{\me' \mA \me}} \exp \left\lbrace \frac{1}{2} \frac{ \left( \me' \mA \mv \right)^{2} }{\me' \mA \me} \right\rbrace . \nonumber
\end{equation}
\normalsize
From this we take $h^* (\me) \propto (\me' \mA \me)^{-\frac{1}{2}}$. On the other hand, one can minimize $I_{\me} ( \mX^{(t+1)}, \mX^{(t)} )$ by maximizing $\exp\{-I_{\me} ( \mX^{(t+1)}, \mX^{(t)} )\}$. Then, choosing
\begin{equation}
h^*(\me) \propto (\me' \mA \me)^{-\frac{1}{2}} \nonumber
\end{equation}
will naturally choose directions with low $I_{\me} ( \mX^{(t+1)}, \mX^{(t)} )$, as can be seen from (\ref{eqn.Ie}).

To sample from $h^*(\me)$, it is also easy to see that
$h^*(\me) \propto \int \pi( \mmu + \tau\me ) d\tau \propto (\me' \mA \me)^{-\frac{1}{2}}$ (the maximum is achieved
when $\mx = \mmu$).  If we simulate $\me_u$ from a Multivariate Normal with precision matrix $\mA$ and centered at the origin, $\pi_0$, and
take $\me = \frac{\me_u}{|| \me_u ||}$, it is clear that $\me$ will have density
$\propto \int \pi_0( \tau\me ) d\tau \propto (\me' \mA \me)^{-\frac{1}{2}}$.  That is, $\me \sim h^*$, and this density
has the whole sphere $\mathbb{S}^n$ as its support and this results in an ergodic chain.

\section{Non-normal objective function}

So far nothing has been achieved (since for sampling from a multivariate normal several samples form basically the same MN distribution are needed!). But indeed, the aim is to produce an efficient sampler when a (local) Multi- variate Normal approximation exists. Note from the previous section that our Optimal Direction Gibbs algorithm only requires knowledge of the precision matrix $\mA$, not of the mean $\mmu$ (to sample the direction from $h^*$). That is, only an (local) approximation to the the log of the objective function will be needed.

\subsection{Truncated Normal objective function}

Sometimes the quantities we want to determine are different from the ones which we are able to measure. If the data measured depends on the quantities we want, then the data contains some information about those quantities. Inverse problems occurs when observed data $ \my $ depend on unknowns $ \mx $ via a measurement process, and we want to recover $ \mx $ from $ \my $. In linear inverse problems, the relationship between $ \my $ and $ \mx $ is given by
\begin{equation*}
\mB \mx = \my ,  
\end{equation*}
where $\mB $ is a linear transformation. The classical inverse problem is to invert the function $ \mB $ to obtain unknowns $ \mx $ in terms of data. Suppose that $\mB_{m \times n} \mx_{n \times 1} = \my_{m \times 1}$ is a linear inverse problem where the matrix $ \mB $ is known. Suppose that $\my | \mx \sim N_{m}(\mB \mx, \mTau)$ with ${\mTau = diag(\frac{1}{\sigma^{2}_{1}}, \ldots,\frac{1}{\sigma^{2}_{m}} )}$, and suppose $ \mx $ has a Truncate Multivariate Normal Distribution with known mean vector $ \mmu$ and precision matrix $\mA$, where $ x_{i} > 0 $ $\forall i $. We are interested in find the posterior distribution of $\mx | \my$. After some algebra, we see that 
\small
\begin{align*}
f( \mx | \my) &\propto \exp \left\{ -\frac{1}{2} ((\mx - \mmu^{*})'\mA^{*}(\mx - \mu^{*}))\right\} \\
& \quad \cdot 1(x_{i}>0)
\end{align*} 
\normalsize
is a Truncate Multivariate Normal Distribution, with precision matrix 
\begin{equation*}
\mA ^{*} = ( \mA + \mB'(m \mTau) \mB ),\end{equation*}
and mean vector 
\begin{equation*} 
\mmu^{*} = ( \mA + \mB'(m\mTau) \mB )^{-1} ( \mmu '\mA + \bar{\my}'(m\mTau) \mB ).
\end{equation*}

We take a Multivariate Normal distribution $\pi$ with precision ($n \times n$) matrix $\mA$ and mean vector
$\mmu = \left( \sqrt{1 \over n}, \ldots , \sqrt{1 \over n} \right) $, but truncated the support to $x_i \geq 0$ (all entries are positive, our objective function is $\propto \pi(\mx)$).
A random direction Gibbs sampling is considered with direction distribution $h^*$ as in
Section~\ref{sec.directions}.  We set the initial point $\mX^{(0)} =  \mmu$ and the burn in is not needed.  The full conditionals are the same as
in Section~\ref{sec.normal} but bound by the positivity constraint.

We consider several dimensions $n = 2, 3, 5,$ $10, 15, 20$.  In each case, we compare with a standard random scan 
(alone the base axis) Gibbs sampling, a pure random direction Gibbs and directions simulated from $h^*$.    
Using the QR decomposition of a $n \times n$ matrix of uniform random entries one obtains a random orthonormal matrix $\boldsymbol{P}$, which
will represent the orthonormal base of eigenvalues.  The eigenvectors of $\mA$, $\lambda_1 , \cdots , \lambda_n$,
represent the precisions on each eigenvalue direction and are set to $\lambda_i = \sigma^{-2}_i$.  The standard deviations
in each principal (eigen) direction are set to
\begin{equation}\label{eqn.sigmas}
\sigma_i = i^{- \alpha \over n}, 
\end{equation}
$\alpha \geq=0$, and $\mA = \boldsymbol{P}' \boldsymbol{\Lambda} \boldsymbol{P}$, where $\boldsymbol{\Lambda} = diag(\lambda_i )$.
These represent decreasing standard deviations
and are increasingly contrasting as $\alpha$ increases.  More contrasting standard deviations result in further correlated distributions.
We consider
$\alpha = 0, 5, 10, 20$.

\begin{figure*}
\begin{center}
\begin{tabular}{c c}
\includegraphics[height=6cm, width=7cm]{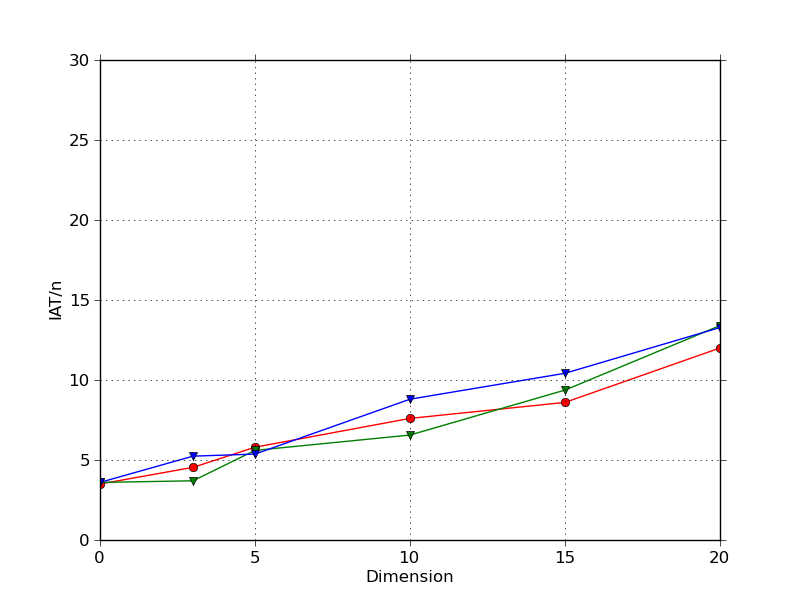} &
\includegraphics[height=6cm, width=7cm]{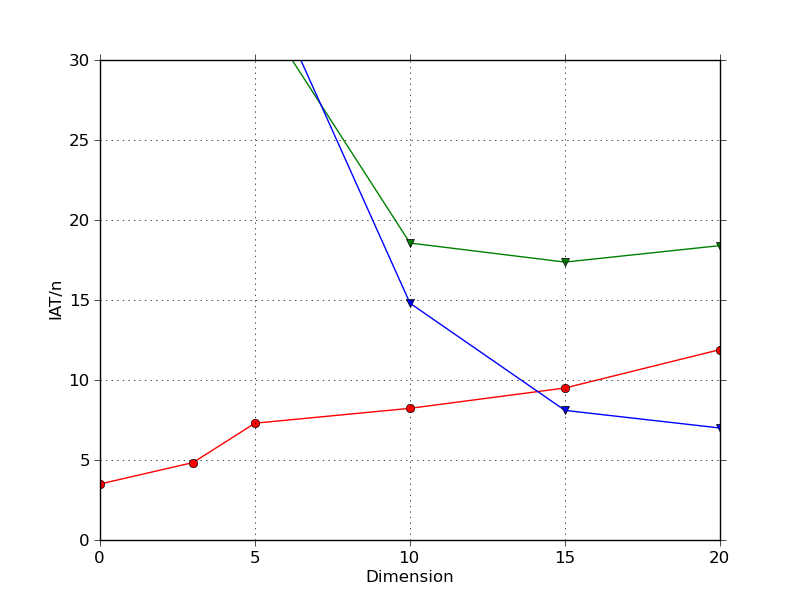} \\
(a) & (b) \\
\includegraphics[height=6cm, width=7cm]{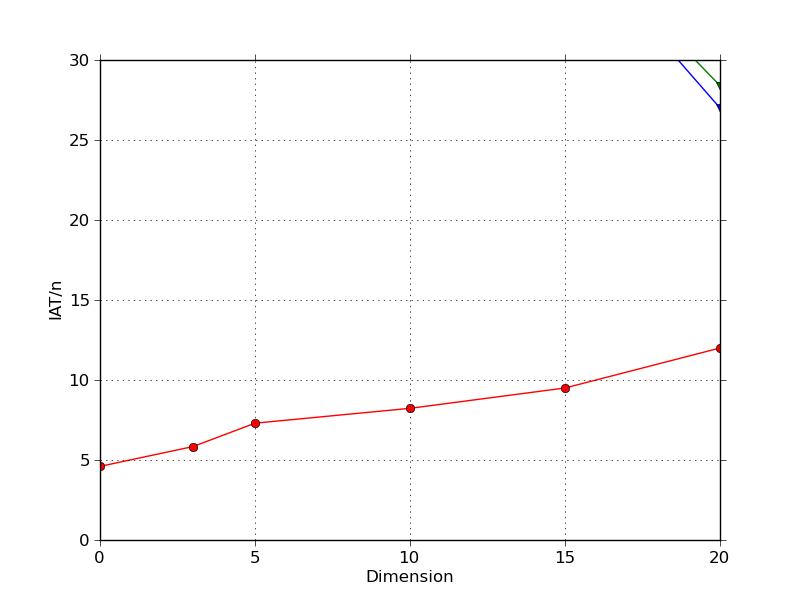} &
\includegraphics[height=6cm, width=7cm]{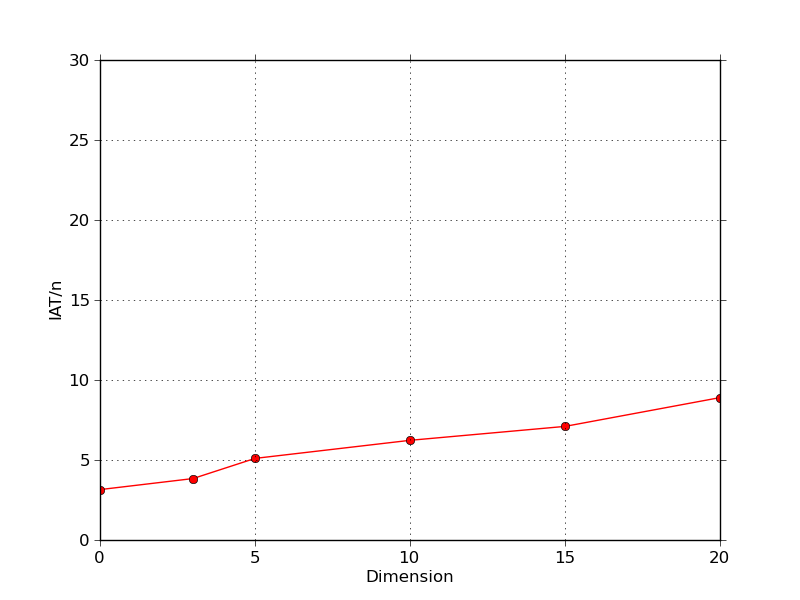} \\
(c) & (d)
\end{tabular}
\caption{\label{fig.testnorm} 
Integrated autocorrelation times divided by the dimension of the MN objective function.
Direction gibbs for random scan (blue), random direction (green) and direction according to $h^*$ (red).  The standard
deviations for the MN are set according to (\ref{eqn.sigmas}) for (a) $\alpha=0$ (independent normals) (b) $\alpha=5$,
(c) $\alpha=10$ and (d) $\alpha=20$.}
\end{center}
\end{figure*}

For each combination of $n$ and $\alpha$ we calculate the Integrated Autocorrelation Time of the resulting
chains.  Results are shown in Figure \ref{fig.testnorm}.  Note that the random scan and random direction Gibbs
worsen in performance as the $\sigma_i$'s are more contrasting, our Optimal Direction Gibbs (ODG) remains with an
IAT/$n \leq 12$.  Note that in this general case, the ODG is a random walk; the resulting IAT's are quite close to the
theoretical optimum for an optimal scaled random walk, as explained in \citep{Roberts:Rosenthal:01}.

\subsection{More general objective functions} \label{other_objective}

Suppose then that $\pi(\mx)$ is the objective distribution of interest with support $\mathcal{X} \subset \mathbb{R}^n$.
Let $\mX \in \mathbb{R}^n$ be a random variable with density $\pi(\mx)$.  Suppose also that for each
$\mx \in   \mathcal{X}$ we have the Hessian $H(\mx)$ for $ - \log \pi(\cdot)$ evaluated at $\mx$.
By completing the squares in the second order taylor approximation to $ - \log \pi(\cdot)$, we assume sufficient smoothness
such that
\small
\begin{equation}
- \log \pi(\my) \approx C(\mx) + \frac{1}{2} (\my - \mmu(\mx))' H(\mx) (\my - \mmu(\mx)), \nonumber
\end{equation}
\normalsize
where $\mmu(\mx) = \mx - H^{-1}(\mx) \nabla(\mx)$ ($\nabla(\mx)$ is the gradient of $- \log \pi(\cdot)$ evaluated at $\mx$).
That is, we have a local MN approximation to $\pi$ with mean vector
$\mmu(\mx)$ and precision matrix $H(\mx)$.  

The proposed algorithm runs as follows.  At $\mX^{(t)} = \mx $ \textit{propose} jumping to 
\begin{equation}\label{eqn.prop}
\mY = \mx + r \me  %= \mX^{(t+1)} 
\end{equation}
where the length $r \in \mathbb{R}$ has distribution $r \sim N\left( -\frac{\me' H(\mx) \mv}{\me' H(\mx) \me}, \me' H(\mx) \me \right)$
and $\mv = \mx - \mmu(\mx) = \mx - \mx + H^{-1}(\mx) \nabla(\mx)$.  This translates to
\begin{align}
r &\sim N\left( -\frac{\me' H(\mx) H^{-1}(\mx) \nabla(\mx)}{\me' H(\mx) \me}, \me' H(\mx) \me \right) \nonumber \\
&= N\left( -\frac{\me' \nabla(\mx)}{\me' H(\mx) \me}, \me' H(\mx) \me \right) . \nonumber
\end{align}
As usual, the proposal $\mY$ will be accepted with a Metropolis-Hastings probability.  In the case that $\pi$ is a MN as in Section~\ref{sec.normal}
it is clear that $\nabla(\mx) = \mA (\mx - \mmu)$, $H(\mx) = \mA$, therefore $\mv = \mA^{-1} \mA (\mx - \mmu)$ and we are back
with the same sampler. The Metropolis-Hasting ratio would be
\begin{equation}
R = \frac{\pi(\my)}{\pi(\mx)} \frac{q(\mx | \my)}{q( \my | \mx )} , \nonumber
\end{equation}
where $q( \cdot | \mx ) = h^*(\me) q_{\me}( \cdot | \mx )$ is the density of $\mY$ in (\ref{eqn.prop}).
In Section~\ref{sec.normal} $q_{\me}( \cdot | \mx )$ is the conditional distribution over the direction $\me$, given $\mx$ (gibbs kernel),
and cancels out with $\pi(\mx)$ ($h^*(\me)$ would cancel out with $h^*(-\me)$ since the Hessian is constant; $H(\mx) = \mA$).
Let $g(\mx) = \log \pi(\mx) $, then
\begin{equation}
\log R = (g(\my) + \log q(\mx | \my)) - (g(\mx) + \log q(\my | \mx)) . \nonumber
\end{equation}
Let
\begin{equation}
\psi_{\mx} ( \me , r) = g(\mx) + \log q(\mx + r \me | \mx) . \nonumber
\end{equation}
Note that $q(\mx + r \me | \mx) = h^*(\me) q_{\me} (\mx + r \me | \mx)$ and therefore
\small
\begin{align}
q(\mx + r \me | \mx) &= K_{H(\mx)} (\me'H(\mx)\me)^{- {1 \over 2}} \frac{(\me'H(\mx)\me)^{1 \over 2}}{\sqrt{2 \pi}} \nonumber \\
& \quad \cdot \exp \left\{ -\frac{\me'H(\mx)\me}{2} \left( r + \frac{\me' \nabla(\mx)}{\me'H(\mx)\me} \right)^2 \right\} , \nonumber
\end{align}
\normalsize
and
\begin{equation}
\psi_{\mx} ( \me , r) = g(\mx) + \log K_{H(\mx)} 
 -\frac{\me'H(\mx)\me}{2} \left( r + \frac{\me' \nabla(\mx)}{\me'H(\mx)\me} \right)^2 , \nonumber
\end{equation}
where
\begin{equation}
K_{H(\mx)}^{-1} = \int_{|| \me || = 1}  (\me'H(\mx)\me)^{- {1 \over 2}} d\me \nonumber
\end{equation}
is the normalizing constant of $h^*$ that now depends on $\mx$.  We then have
\begin{equation}\label{eqn.R}
R( \mx, \me, r) = \exp \{ \psi_{\mx + r \me} ( -\me , r) - \psi_{\mx} ( \me , r) \}.
\end{equation}
Therefore, the probability of accepting a jump from $\mx$ to $\my = \mx + r \me$ should be $\min \{ 1 , R( \mx, \me, r) \}$.
However, it is not possible to obtain $K_{H(\mx)}$ analytically.  For dimension $n=2$ we calculate $K_{H(\mx)}$ numerically, only for comparison purposes.

\indent An alternative distribution of directions would be the following. Assuming that $\mA = H(\mx)$ is positive defined, we can use its eigenvalues and eigenvectors. We will take the directions $ \me $ as the eigenvectors of matrix $ H(\mx) $, so $\me \in \{ \me_{1}, \me_{2}, \ldots , \me_{n} \} $. The $ i $-th direction will be selected with probability proportional to $ \lambda_{i}^{-1} $, where $ \lambda_{i} $ is the eigenvalue corresponding to the $ i $-th eigenvector , $i = 1,2,\ldots ,n$ . Then $ h_{1}( \me_{i} ) = \left( k \lambda_{i} \right) ^{-1} $, where $ k = \sum _{i=1}^{n} \lambda_{i}^{-1}$. By doing this the distribution of $ \mathbf{e} $ becomes discrete. Moreover, we avoid the problem we have with de normalization constant $ K_{H(\me)} $ and the implementation can be done for large values of dimension $ n $.

It Iis easy to see that direction $ \mathbf{e}_{1} $, corresponding to the lowest eigenvalue $ \lambda_{n} $ of $ H(\mx) $, is optimal indeed. Note that
\begin{align*}
\min _{\Vert \me \Vert = 1} I_{\me} \left( \mX ^{(t+1)}, \mX^{(t)} \right) & = 
\min _{\Vert \me \Vert = 1} \left\lbrace C + \dfrac{1}{2} \log \left( \me ' H(\mx) \me \right) \right\rbrace \\
& = C + \dfrac{1}{2} \log \left( \min _{\Vert \me \Vert = 1} \left\lbrace \me ' H(\mx) \me \right\rbrace \right) =
C + \dfrac{1}{2} \log \lambda_{1}.
\end{align*}
\normalsize
The minimum is reached when $ \me = \me_{1}$, the eigenvector associated to $ \lambda_{1} $.

\subsection{Example: The skew-normal distribution}

To test the performance of our algorithm we will use as objective distribution a variant of the skew-normal distribution.  We choose this distribution due its simplicity and versatility, plus the fact that it allows us to calculate analytically the gradient and the Hessian of $ - \log \pi $.

The multivariate skew-normal probability density function is given by
\begin{equation}
f( \my ) = 2 \phi_{n}( \my - \boldsymbol{\xi} ; \mSigma ) \Phi ( \malpha ' ( \my - \boldsymbol{\xi} ) ) , \nonumber
\end{equation}
where $ \phi_{n}( \cdot ; \mSigma )  $ is the multivariate normal probability density function with mean vector $ \mathbf{0} $ and variance-covariance matrix $ \mSigma $, $ \Phi ( \cdot )  $ is the standard univariate normal cumulative distribution function, $ \boldsymbol{\xi} $ is a localization parameter and $ \malpha $ is a shape parameter   \citep{Azzalini:Adelchi:05}.

$ \Phi ( \cdot ) $ is called \textit{perturbation} function and may be the cumulative distribution function of any univariate symmetric distribution around $ 0 $. In our case, we are going to use the Logistic cumulative distribution function with mean $ 0 $ and scale parameter $ s = \frac{ \sqrt{3} }{ \pi} $, which is

\begin{equation}
G(x) = \dfrac{1}{ 1 + \exp \left\lbrace - \pi x / \sqrt{3} \right\rbrace }. \nonumber
\end{equation}
\begin{figure*}
\begin{center}
\begin{tabular}{c c}
\includegraphics[height=5cm, width=6.9cm]{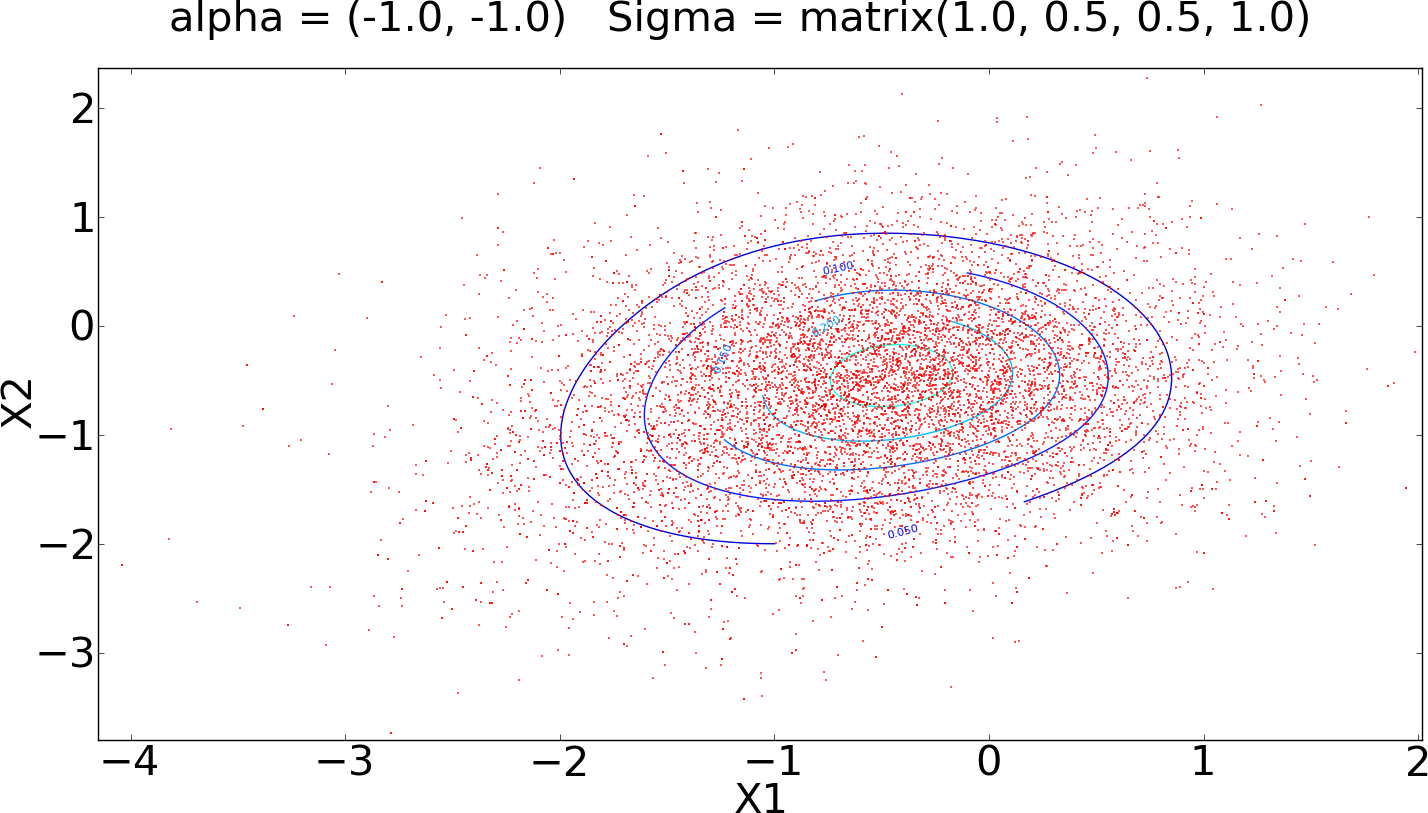} &
\includegraphics[height=5cm, width=6.9cm]{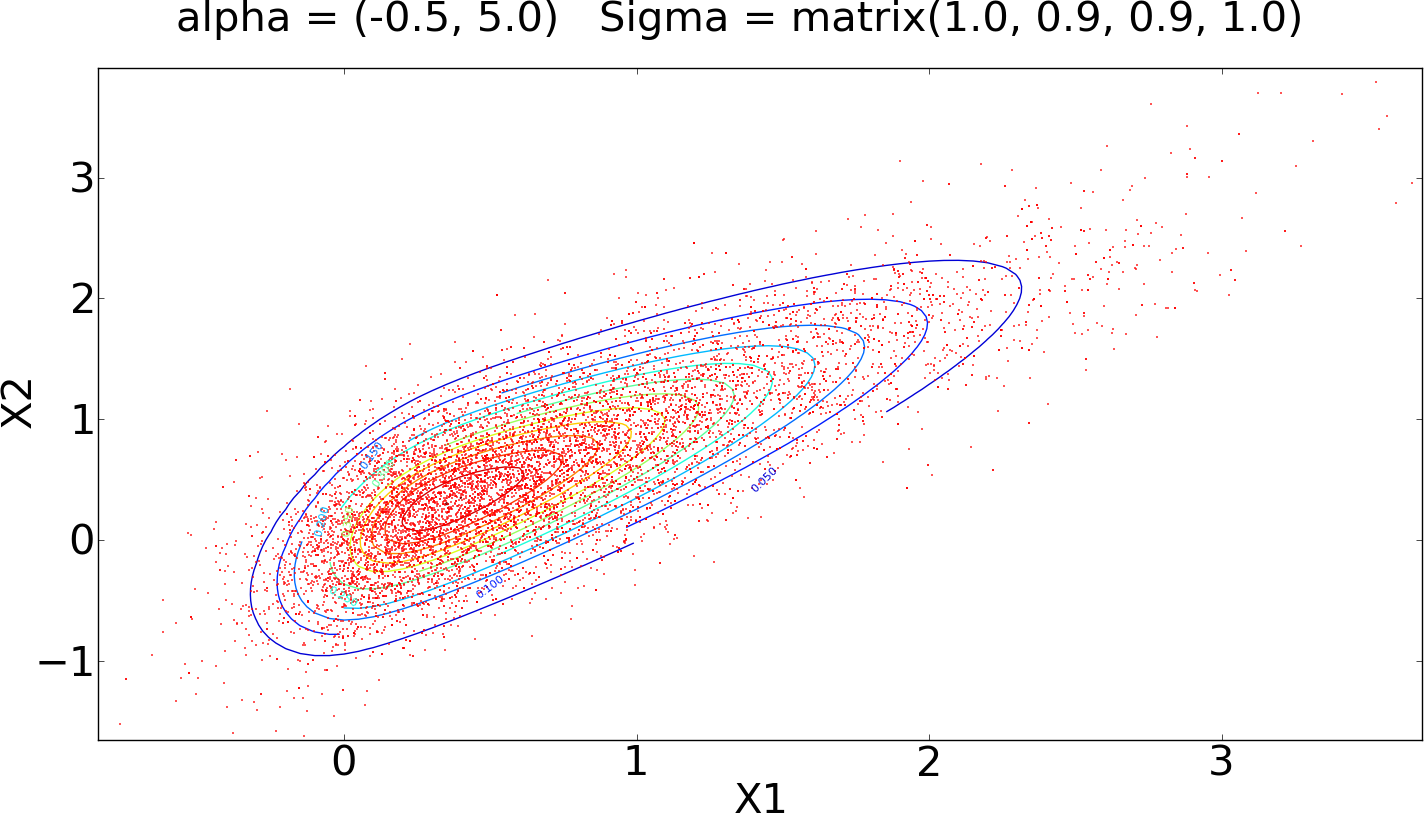} \\
a) & b) \\
   &    \\
\includegraphics[height=5cm, width=6.9cm]{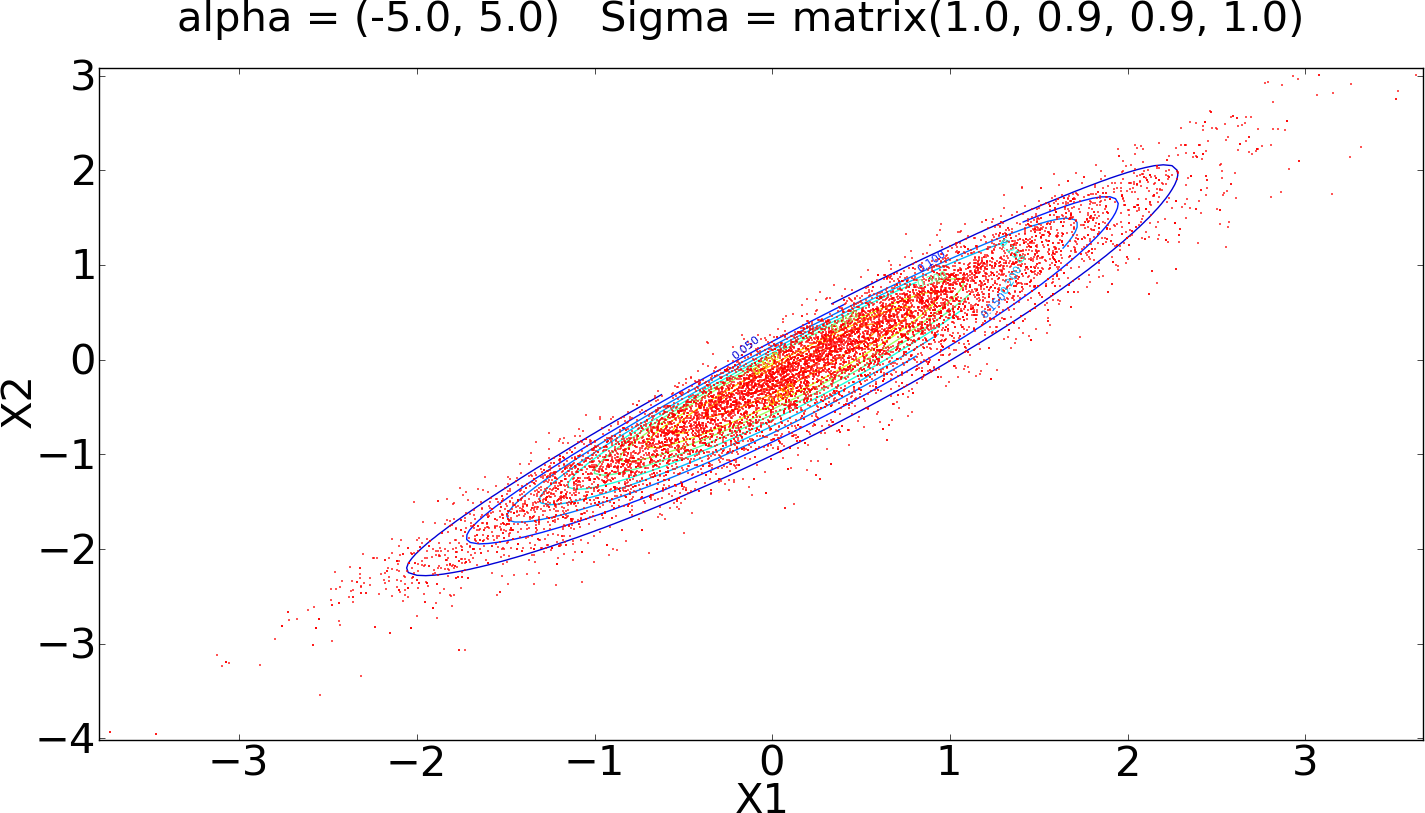} &
\includegraphics[height=5cm, width=6.9cm]{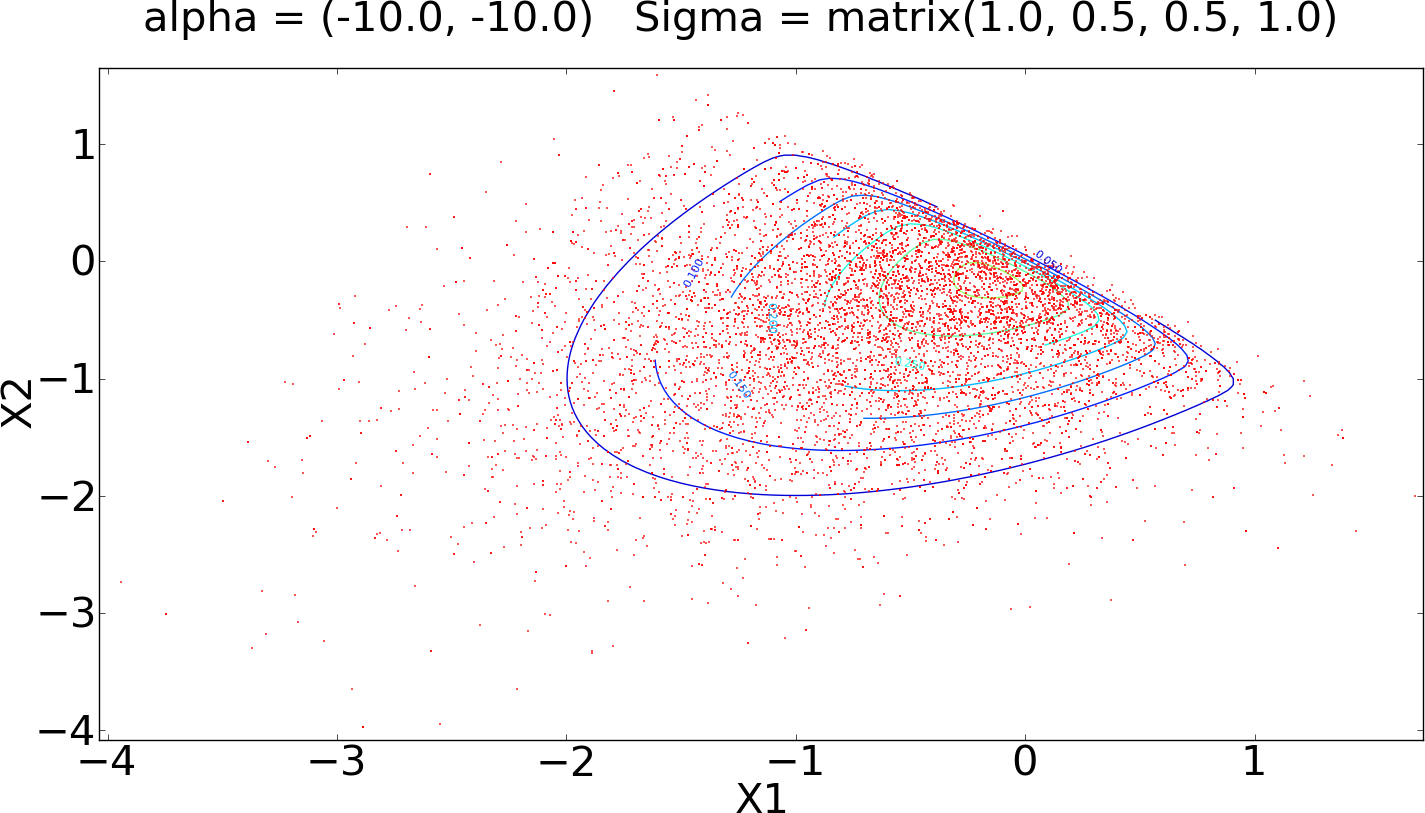} \\
c) & d) \\
\end{tabular}
\caption{\label{fig.h_skew_normal} Simulations obtained with the Optimal Direction Gibbs algorithm using distribution $ h* $. }
\end{center}
\end{figure*}

\begin{table*}
\begin{center}
\begin{tabular}{|c|c|c|c|c|}
\hline    & $ \malpha $        & $ \mSigma $              & IAT          & Acceptance rate \\  
\hline a) & $ (- 1.0, - 1.0) $ & $ (1.0, 0.5, 0.5, 1.0) $ & $ 4.039870 $ & $ 0.8712 $ \\ 
\hline b) & $ (- 0.5,   5.0) $ & $ (1.0, 0.9, 0.9, 1.0) $ & $ 6.944900 $ & $ 0.7693 $ \\ 
\hline c) & $ (- 5.0,   5.0) $ & $ (1.0, 0.9, 0.9, 1.0) $ & $ 3.909896 $ & $ 0.8485 $ \\ 
\hline d) & $ (-10.0, -10.0) $ & $ (1.0, 0.5, 0.5, 1.0) $ & $ 7.629253 $ & $ 0.6892 $ \\
\hline 
\end{tabular}  
\caption{\label{tableh} Acceptance rate and IAT for simulations obtained with the Optimal Direction Gibbs algorithm using distribution $ h* $.}
\end{center} 
\end{table*}

Then, the density function of our objective distribution is given by
\begin{equation}
\pi ( \mx ) =  2 \left( \dfrac{\vert A \vert}{(2 \pi)^{n}} \right) ^{1/2} \exp \left\lbrace - \dfrac{1}{2} \mx ' \mA \mx \right\rbrace
 \dfrac{1}{ 1 + \exp \left\lbrace -\pi \malpha ' \mx / \sqrt{3} \right\rbrace }, \nonumber
\end{equation}
where $ \mA = \mSigma^{-1} $.
In order to implement the algorithm we need to calculate the gradient and Hessian of $ - \log \pi ( \mx ) $. It is easy to see that
\begin{equation}
\nabla ( \mx ) = \mA \mx - \left( 1- G ( \malpha ' \mx ) \right) \malpha , \nonumber
\end{equation}
and
\begin{eqnarray}
H( \mx ) = A + g ( \malpha ' \mx ) \malpha \malpha ' , \nonumber
\end{eqnarray}
where $ g( \cdot) $ is the Logistic probability density function.

To test the algorithm we consider the case $ n=2 $.  Figure \ref{fig.h_skew_normal} shows the contour plots of the objective distribution for different values of $ \mSigma $ and $ \malpha $. It also shows the simulations obtained using the algorithm with distribution $ h^{*} $, after $ 10,000 $ iterations.  We set the initial point $ \mX ^{(0)} =  \mathbf{0} $ and the burn in is not needed. Table \ref{tableh} shows the acceptance rate and IAT for each example. Note that the value of the IAT is small and the acceptance rate is large in all cases.  Both measures indicate that the performance of the algorithm is quite good. Is important to mention that a conventional Gibbs sampler cannot be used for this example since the full conditional distributions do not have any close form.  Our scheme has not that kind of difficulties.

Unfortunately, the implementation is not as simple for dimension $ n \geq 3 $ since we need
to calculate the normalization constant $ K_ {H (\mx)} $ numerically. 

\begin{figure*}
\begin{center}
\begin{tabular}{c c}
\includegraphics[height=5cm, width=6.9cm]{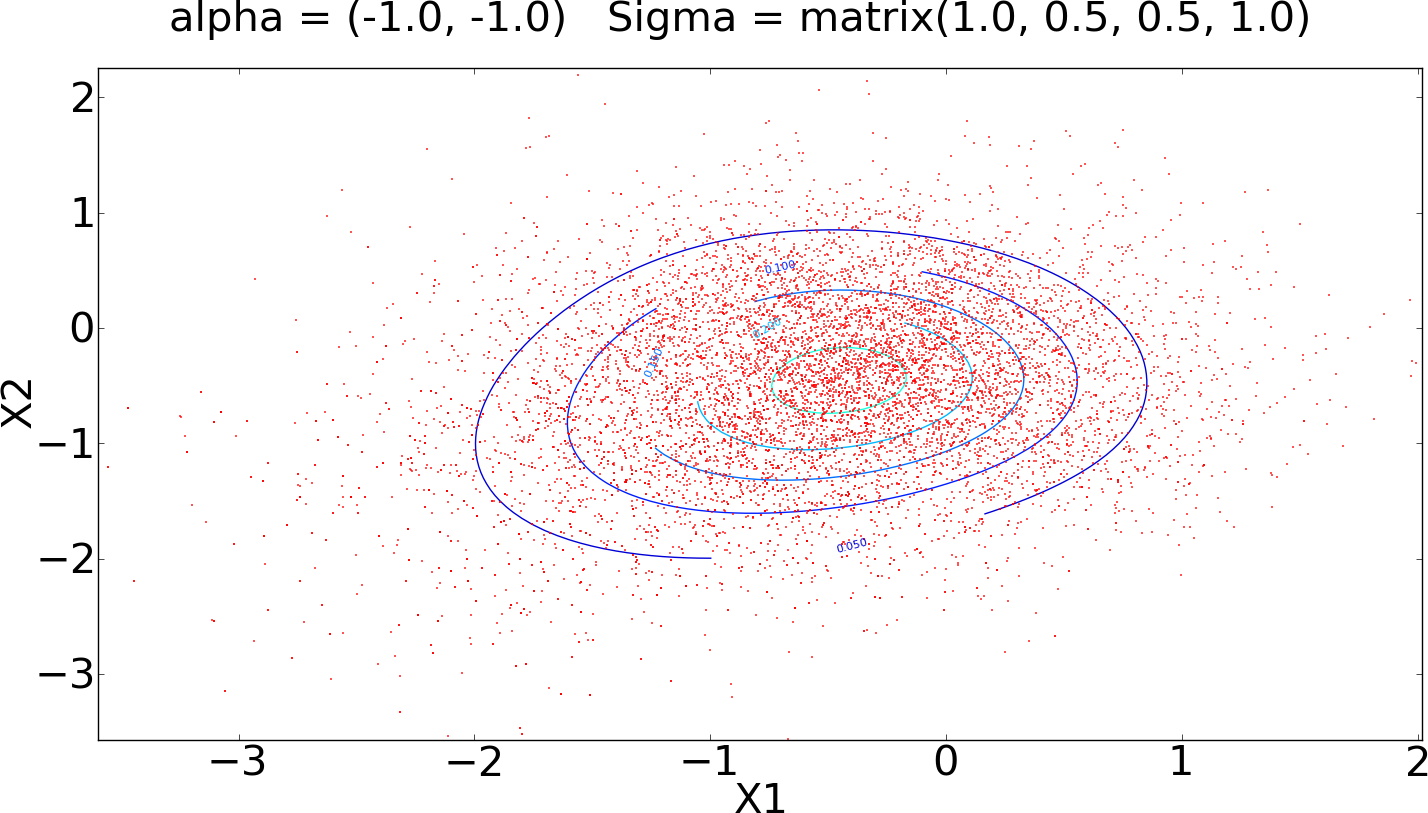} &
\includegraphics[height=5cm, width=6.9cm]{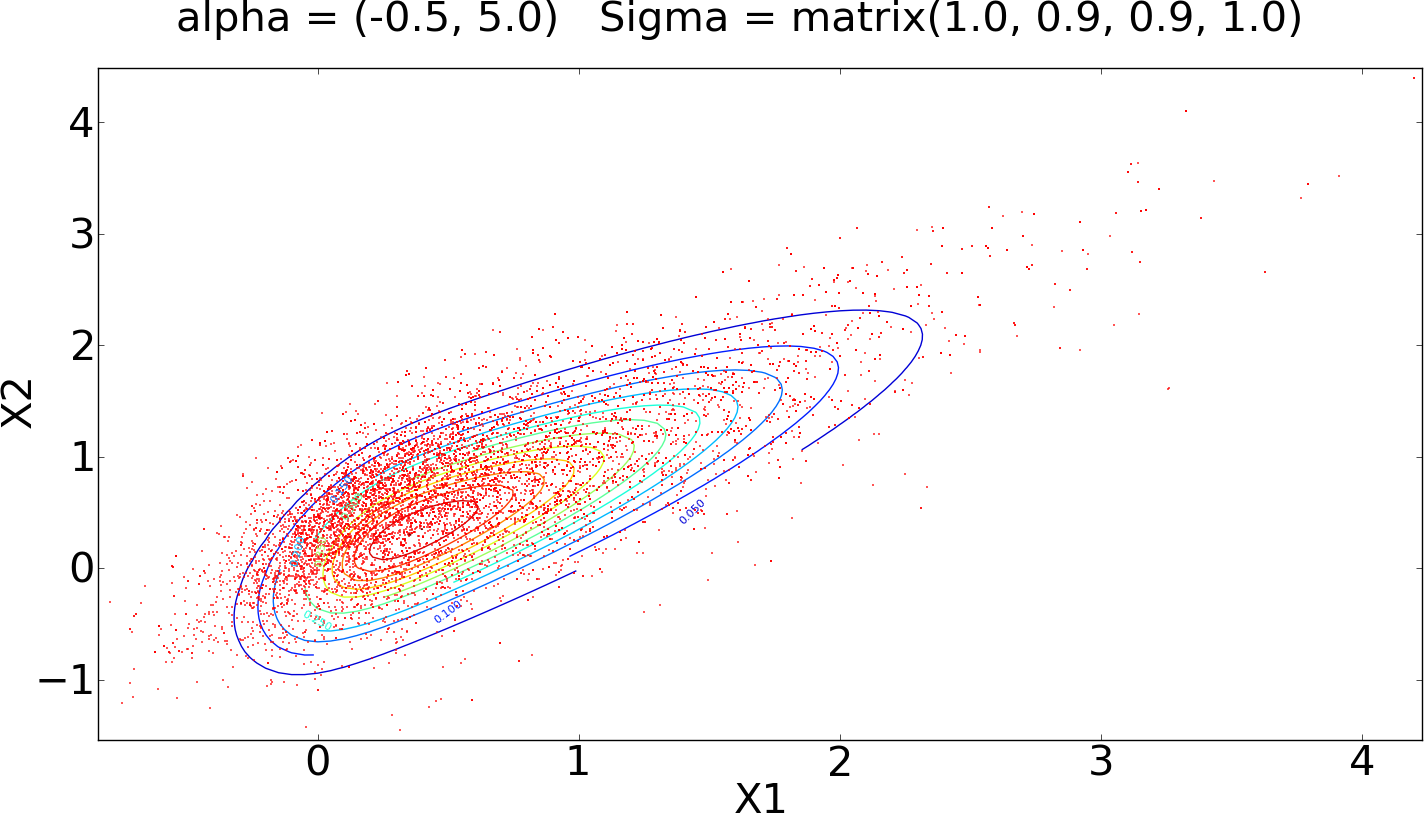} \\
a) & b) \\
   &    \\
\includegraphics[height=5cm, width=6.9cm]{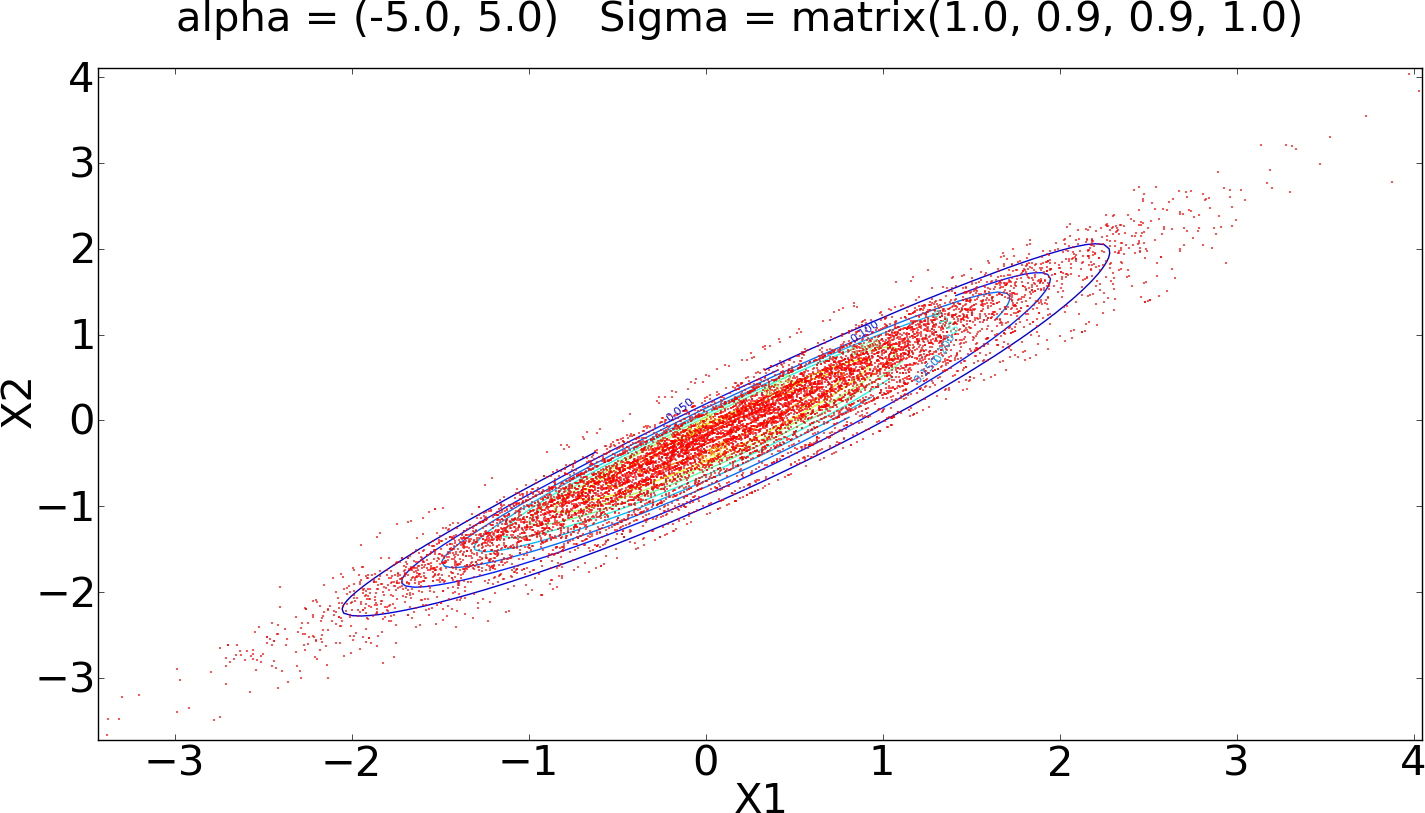} &
\includegraphics[height=5cm, width=6.9cm]{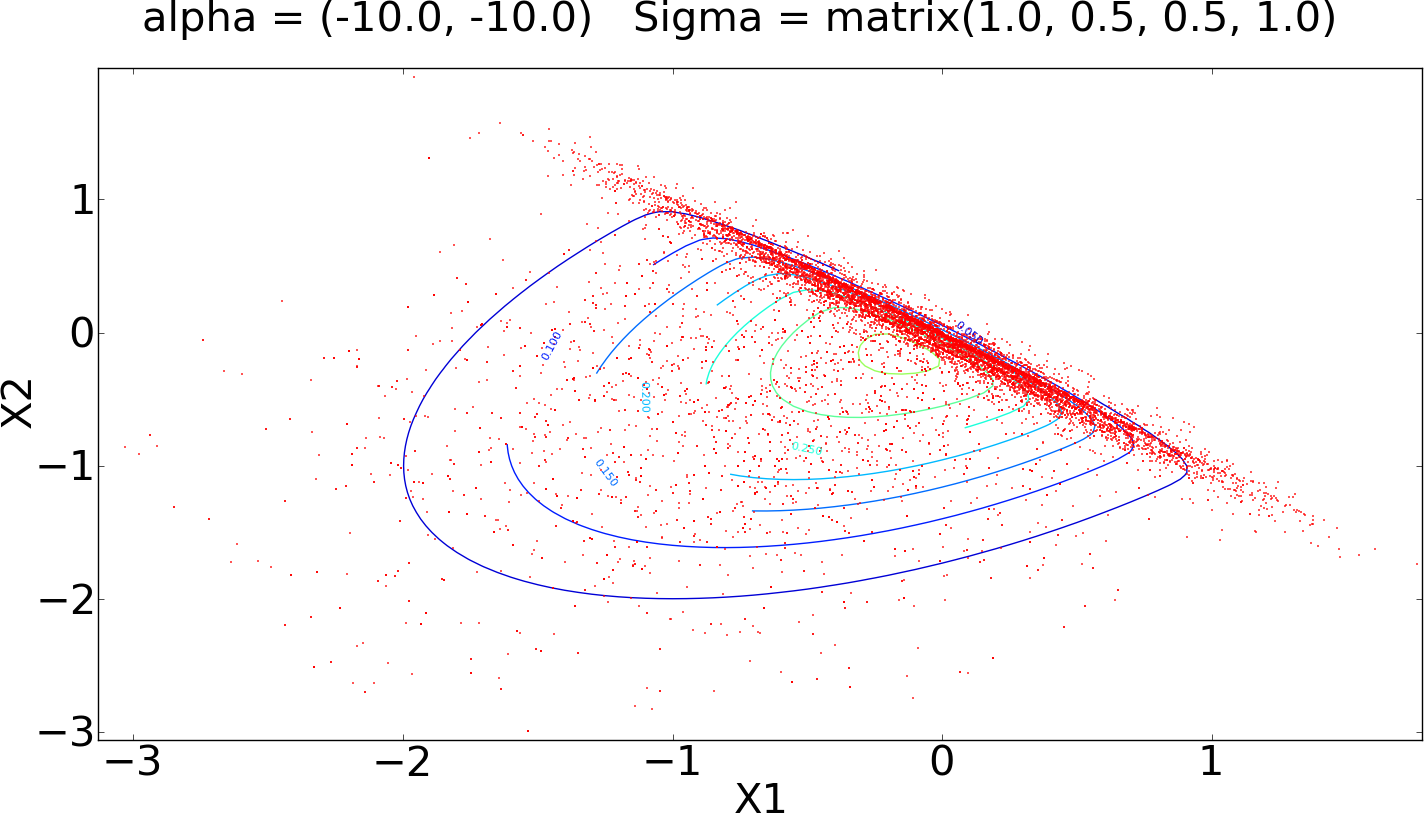} \\
c) & d) \\
\end{tabular}
\caption{\label{fig.h1_skew_normal} Simulations obtained with the Optimal Direction Gibbs algorithm using distribution $ h_{1} $. }
\end{center}
\end{figure*}

In Subsection \ref{other_objective} we proposed an alternative direction distribution $ h_{1} $ to avoid calculating $ K_ {H (\mx)} $.
We test the algorithm using this distribution on the same examples. Figure \ref{fig.h1_skew_normal} shows some very interesting results.  Simulations in examples a) and b) are well distributed over the region of interest. However, in examples c) and d) simulations are concentrated in a specific region. This results from the fact that the eigenvalues of $ H( \mx ) $ are very contrasting, one is much larger than the other, and consequently, simulations are obtained in similar directions constantly.

This suggests that the scheme works well in simple cases, but it becomes inefficient against more skewed distributions.
However, we can still modify $ h_{1} $. We take the directions $ \me $ as eigenvectors of $ H( \mx ) $, but the probability of choosing $ \me_{i} $ will now be proportional to $ \left( \lambda_{i} \right) ^{-b} $, where $ b \sim Beta (1, 9) $, this is $ h_{2} ( \me _{i} ) \propto \left( \lambda _{i} \right) ^{-b} $. By doing this we allow the chance that probabilities become more balanced in regions where there is a very dominant eigenvalue.

\begin{figure*}
\begin{center}
\begin{tabular}{cc}
\includegraphics[height=5cm, width=6.9cm]{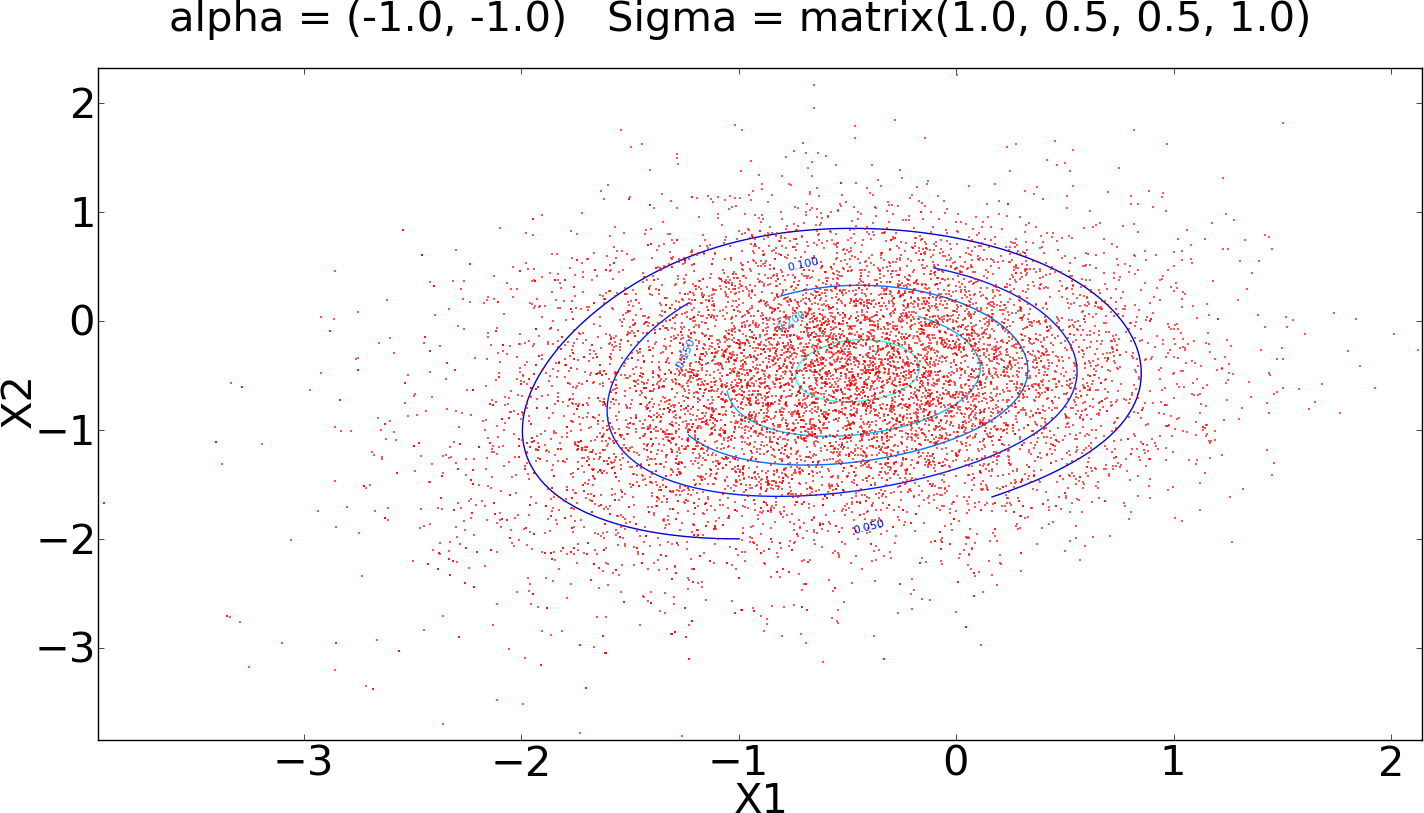} &
\includegraphics[height=5cm, width=6.9cm]{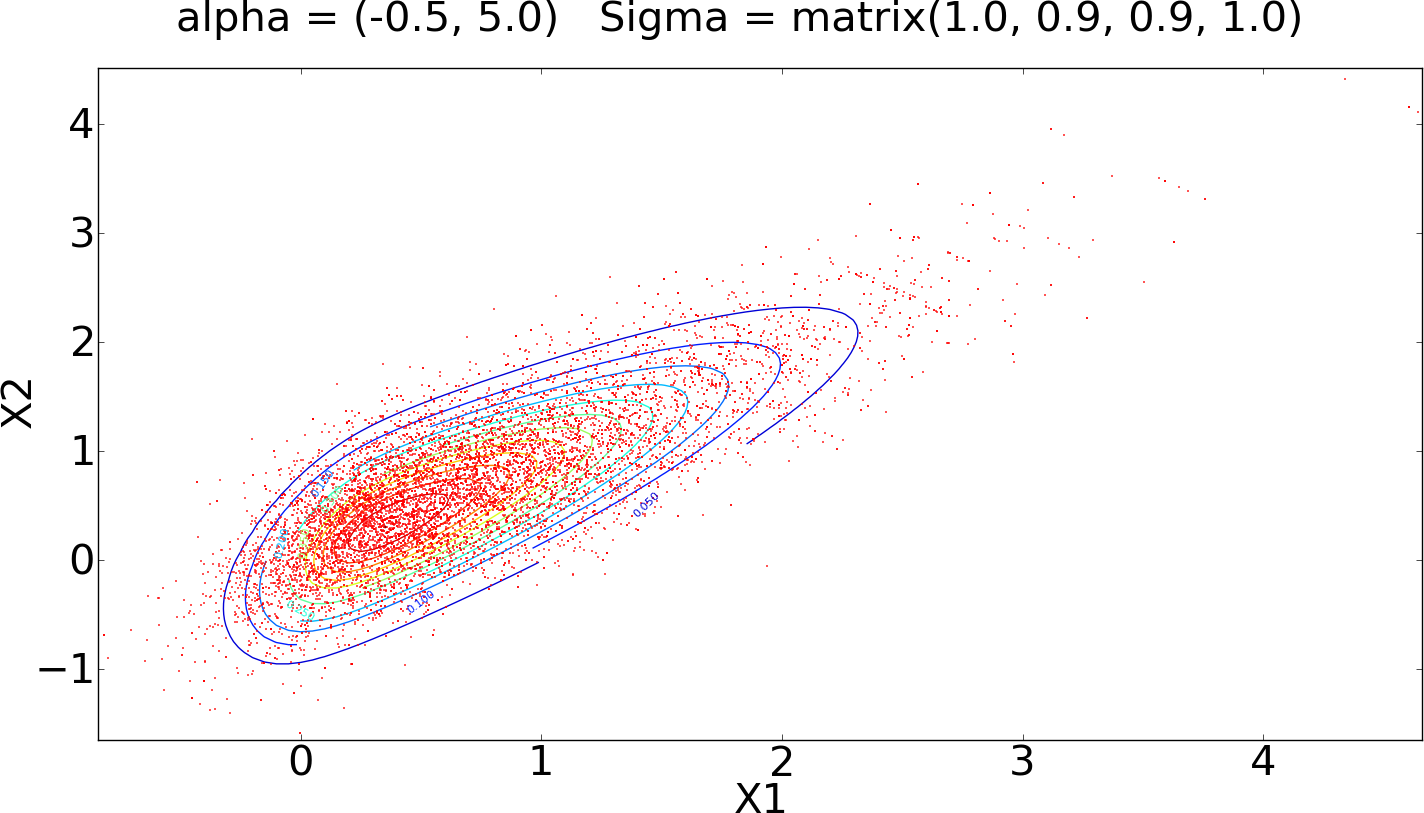} \\
a) & b) \\
   &    \\
\includegraphics[height=5cm, width=6.9cm]{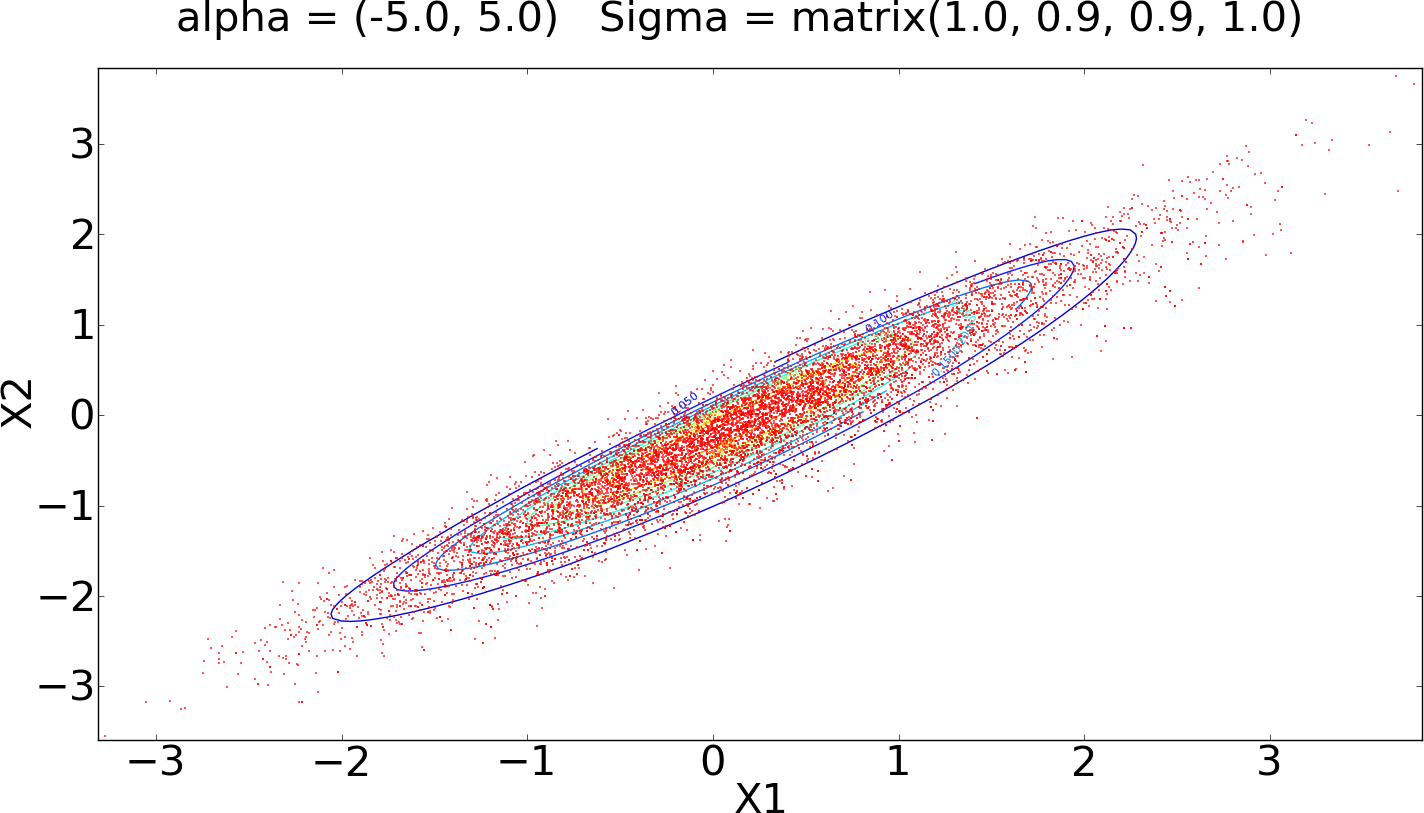} &
\includegraphics[height=5cm, width=6.9cm]{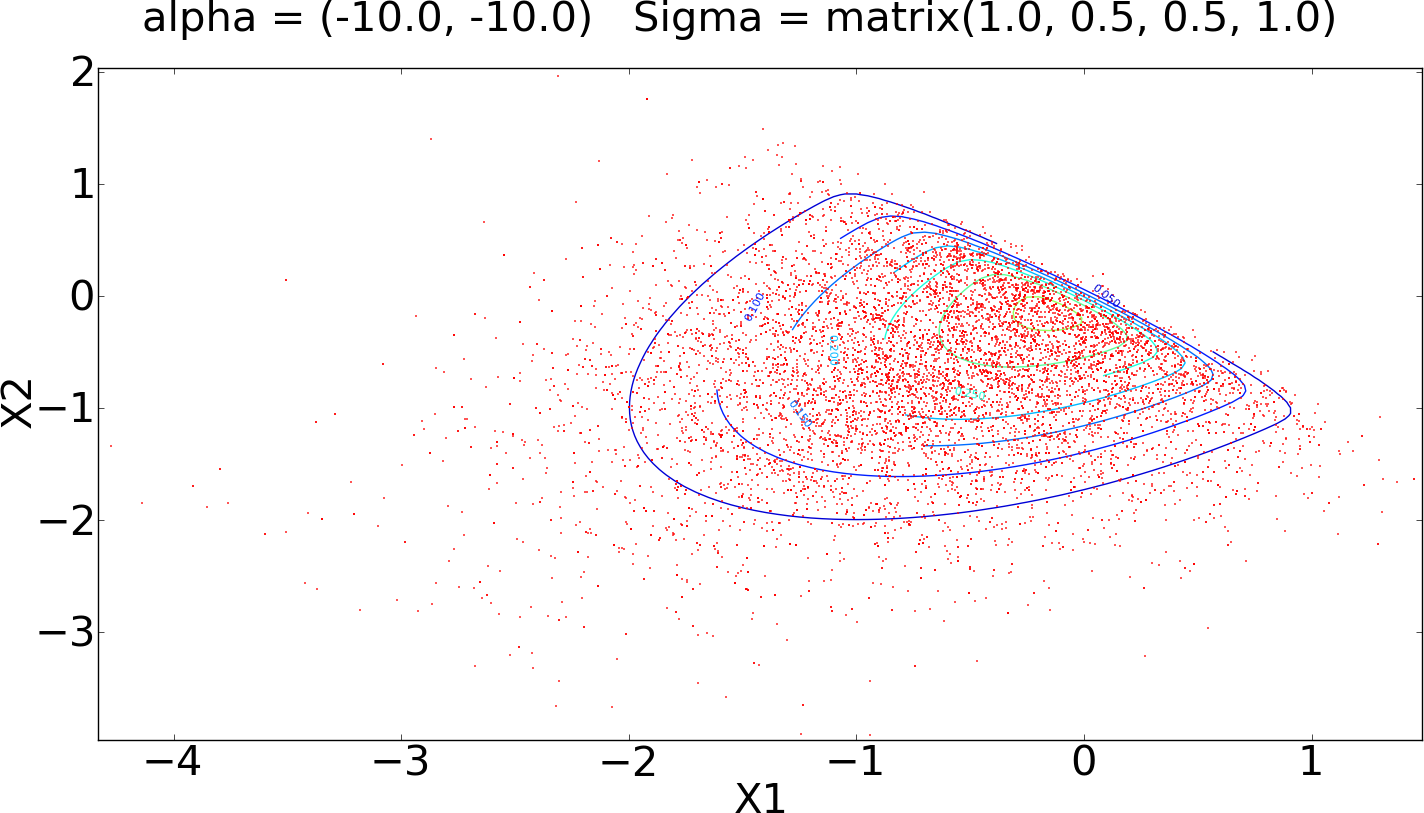} \\
c) & d) \\
\end{tabular}
\caption{\label{fig.h2_skew_normal} Simulations obtained with the Optimal Direction Gibbs algorithm using distribution $ h_{2} $. }
\end{center}
\end{figure*}

Figure \ref{fig.h2_skew_normal} shows the behaviour of the resulting algorithm. Examples a) and b) show a good performance as before. On the other hand, examples c) and d), which we identified as difficult above, seem to have been corrected. Table \ref{tableeigennew} shows the acceptance rate and IAT. Here we see good results. Acceptance rates remain high and do not represent a problem. The IAT is slightly higher than in Table \ref{tableh}, except by example c); however, the difference is not quite significant.

\begin{table*}
\begin{center}
\begin{tabular}{|c|c|c|c|c|}
\hline    & $ \malpha $        & $ \mSigma $              & IAT          & Acceptance rate \\
\hline a) & $ (- 1.0, - 1.0) $ & $ (1.0, 0.5, 0.5, 1.0) $ & $ 4.497301 $ & $ 0.8793 $ \\ 
\hline b) & $ (- 0.5,   5.0) $ & $ (1.0, 0.9, 0.9, 1.0) $ & $ 7.844630 $ & $ 0.7534 $ \\ 
\hline c) & $ (- 5.0,   5.0) $ & $ (1.0, 0.9, 0.9, 1.0) $ & $ 2.705416 $ & $ 0.8809 $ \\
\hline d) & $ (-10.0, -10.0) $ & $ (1.0, 0.5, 0.5, 1.0) $ & $ 8.821470 $ & $ 0.7327 $ \\
\hline 
\end{tabular}  
\caption{\label{tableeigennew} Acceptance rate and IAT for simulations obtained with the Optimal Direction Gibbs algorithm using distribution $ h_{2} $. }
\end{center}
\end{table*}

Direction distribution $ h_{2} $ seems a good alternative for our optimal direction distribution since it shows good performance and the IAT values are very similar to those obtained with our original scheme. The parameters of the Beta distribution were selected intuitively. In fact, those parameters allows us to change distribution $ h_{2} $ in order to get better results.

\section{Discussion}

Our Optimal Direction Gibbs sampler presents interesting characteristics in examples where either a conventional gibbs sampler
is impossible to implement or o ther MCMC methods (eg. a Random Walk Metropolis-Hastings) would be vey difficult to tune.
The truncated normal example is of great relevance in the field of inverse problems and we have also worked with strongly
skewed distributions with contrasting scales, with promising results in all cases.

\section{Acknowledgements}

DAPR and MSC thank CONACyT for a MSc scholarship since part of this work was conducted while finishing their MSc studies at CIMAT.
Part of JAC work was founded by CONACyT grant 128477. 

\bibliography{paper_odg}

\end{document}